\documentclass[aps,prd,floats,floatfix,showpacs,twoside,preprintnumbers,superscriptaddress,nofootinbib]{revtex4}
\usepackage[dvips]{graphicx,pstricks}
\usepackage{amsmath}
\usepackage{amssymb}
\usepackage{epsfig}
\usepackage{color}

\newcommand{\be}{\begin{equation}}
\newcommand{\ee}{\end{equation}}
\newcommand{\ba}{\begin{eqnarray}}
\newcommand{\ea}{\end{eqnarray}}

\def\roughly#1{\mathrel{\raise.3ex\hbox{$#1$\kern-.75em%
\lower1ex\hbox{$\sim$}}}}

\begin{document}

\date{\today}

\title{Thermodynamic instabilities in dynamical quark models with complex 
conjugate mass poles}

\author{S.~Beni\' c\footnote{sanjinb@phy.hr}}
\affiliation{Physics Department, Faculty of Science, University of Zagreb, 
Zagreb 10000, Croatia}

\author{D.~Blaschke\footnote{blaschke@ift.uni.wroc.pl}}
\affiliation{Institut Fizyki Teoretycznej, Uniwersytet Wroc{\l}awski,
50-204 Wroc{\l}aw, Poland}
\affiliation{Bogoliubov Laboratory for Theoretical Physics, JINR Dubna,
141980 Dubna, Russian Federation }

\author{M.~Buballa\footnote{michael.buballa@physik.tu-darmstadt.de}}
\affiliation{Institut f\"ur Kernphysik (Theoriezentrum), 
Technische Universit\"at Darmstadt\\
	D--64289 Darmstadt, Germany}
\begin{abstract}
We show that the CJT thermodynamic potential of dynamical quark models
with a quark propagator represented by complex conjugate mass poles
inevitably exhibits thermodynamic instabilities. 
We find that the minimal coupling of the quark sector to a Polyakov loop 
potential can strongly suppress but not completely remove such instabilities. 
This general effect is explicitly demonstrated in the
framework of a covariant, chirally symmetric, effective quark model.
\end{abstract}
\maketitle

\section{Introduction}

The fate of hadronic matter in extreme environments, e.g., in
the interior of compact stars or in the early universe,
remains one of the most
interesting unanswered questions today. 
Heavy-ion experiments performed at the Relativistic Heavy-Ion
Collider (RHIC) at the Brookhaven National Laboratory
or at the Large Hadron Collider (LHC) at CERN Geneva
show that at sufficiently high temperature or density,
hadronic matter undergoes a phase
transition by dissolving into its constituents: quarks and gluons.
Details of this transition are encoded in the phase diagram of
quantum chromodynamics (QCD).

Due to its non-perturbative nature at low temperatures
and densities, QCD is best studied on the lattice in this regime. 
Results for the QCD equation of state (EoS) have recently become available 
also at physical quark masses and were extrapolated to the continuum
\cite{Aoki:2009sc,Cheng:2009zi,Borsanyi:2010bp,
Borsanyi:2010cj,Bazavov:2010sb,Bazavov:2010bx,
Bazavov:2010pg,Soldner:2010xk}. 
However, at finite densities lattice simulations are still limited
due to the known sign problem.

In this situation, effective models of QCD serve to
interpret and also extrapolate lattice results. 
To be realistic, such a model must dynamically 
break chiral symmetry and
confine the colored degrees of freedom.
In particular, quark and gluon propagators
should strongly differ from their high-energy counterparts.
A suitable continuum, non-perturbative and covariant
approach is provided by the Dyson-Schwinger equations (DSE) 
(for reviews see e. g. 
\cite{Roberts:2000aa,Alkofer:2000wg,Fischer:2006ub}) and
its descendants, the non-local
chiral quark models 
\cite{Plant:1997jr,General:2000zx,GomezDumm:2001fz,GomezDumm:2004sr}
(see also \cite{Blaschke:1999ab,Blaschke:2000gd,Blaschke:2007ce}),
most recently augmented by the Polyakov loop
(PL) \cite{Blaschke:2007np,Contrera:2007wu,Hell:2008cc,Noguera:2008cm,
Hell:2009by,Contrera:2010kz,Horvatic:2010md,Radzhabov:2010dd}.

A non-perturbative quark propagator is a solution of its DSE,
within the appropriate symmetry-preserving truncation scheme.
The most crucial behavior of these continuum studies is a
strong infrared running of the quark dynamical mass, 
which is to be interpreted as dynamical chiral
symmetry breaking (DChSB),
and of the wave function renormalization. 
Confinement can be realized through the absence of 
poles of the quark propagator at real time-like four-momenta
by the criteria
of positivity violation \cite{Roberts:2000aa,Alkofer:2000wg}.
Indeed, it is a common feature that, due to strong dynamics,
the poles are pushed deep
into the complex four-momentum region
\cite{Krein:1990sf,Burden:1991gd,
Stainsby:1992hy,Burden:1997ja,Gribov:1999ui,Alkofer:2003jj}.
The most simple resulting structure is a series of quartets 
of complex conjugate mass poles (CCMPs).

For a successful phenomenological study it is sufficient to model
the strong interactions via an effective gluon propagator to be
fixed by the infrared observables in the vacuum.
Enhancement of the interaction
in the infrared pushes the quark poles away from
the real axis. 
Another possible realization is the absence of propagator poles in the entire 
complex plane which can be realized either by the presence of cuts instead 
of poles \cite{Buballa:1992sz} or by the absence of both 
when the quark propagator is obtained as an entire function 
\cite{Munczek:1983dx,Efimov:1995uz}.
A non-constant entire function must have a singularity at infinity.
In order for such a quark state to become deconfined and to restore its 
approximate chiral symmetry, e.g., at high temperature, the gluon sector of 
the theory must be restructured in the transition region 
in order to allow for the appropriate changes in the analytic properties 
of the propagators such as the appearance of quasiparticle poles.
But in that case, the usual strategy is no longer applicable: 
to predict the behavior at finite temperature and chemical potential 
from a 
straightforward generalization using the Matsubara formalism without changing
the analytic properties which were adjusted by constraints from vacuum
observables.
A modification of this strategy lies beyond the scope of the present study.

A CCMP structure in the quark propagator is sufficient to ensure
violation of reflection positivity 
\cite{Burden:1997ja,Alkofer:2003jj,Bhagwat:2003vw},
and as such provides a useful form to fit the lattice quark propagator
\cite{Alkofer:2003jj,Bhagwat:2003vw}.
This led to applications at
finite quark chemical potential \cite{Chen:2008zr},
or for parton distribution functions \cite{Tiburzi:2003ja}.
On the other hand, QCD bound states are affected by this structure 
\cite{Bhagwat:2002tx}, e.g.,
a sufficiently heavy meson state (typically of the order of $1$ GeV)
has unphysical $\bar{q}q$ thresholds
\cite{Scarpettini:2003fj} if a n\"aive analytic continuation
to the mass pole of the bound state is used (see \cite{Plant:1997jr}
for a suggestion how this problem could be circumvented by a more
elaborate analytic continuation).
  
In this work we want to further investigate 
the properties of models with CCMPs
by concentrating on the finite temperature and finite
quark chemical potential EoS.
We find that the pressure of quark
matter leads to an unsatisfactory scenario;
the EoS exhibits oscillations in temperature
which are in turn a consequence of the imaginary part of the CCMPs.
We provide analytic insight into the nature of these oscillations, and
suggest a partial solution to this problem by coupling the system
to the Polyakov loop (PL).

This paper is organized as follows. 
In Section II, we present our arguments in a general form by
postulating a CCMP parametrization of the quark propagator for which 
we then obtain the kinetic part of the QCD partition
function in the quark sector.
This result elucidates that the presence of complex conjugated mass poles 
in the quark propagator entails thermodynamic instabilities.
We introduce the PL variable in the partition function 
and show that this step very effectively suppresses the instability.
A separate analysis is performed for the EoS at zero temperature where for
the CCMP parametrization the quark number density and the pressure can be 
obtained in closed form.
In Section III we present the example of dynamical quark
models with chirally invariant nonlocal interaction, including 
explicit numerical results and their discussion. 
In Section IV we give the summary and conclusions of our study.

\section{Thermodynamics in the CCMP representation}
\label{sec:EoS}

The (unrenormalized) thermodynamic potential for the quark sector of QCD 
can be given in the form of the
Cornwall-Jackiw-Tomboulis (CJT) effective action 
\cite{Cornwall:1974vz} as
\be
\Omega(T,\mu) = \Gamma[S]= 
- \mathrm{Tr}\mathrm{Log}(S^{-1})+\mathrm{Tr}[\Sigma S]+
 \Psi[S] ~ ,
\label{CJT}
\ee
where $S^{-1}$ represents the inverse of the full quark propagator
in Euclidean space,
\be
S^{-1}(\tilde{p}_n)=i(\boldsymbol{\gamma}\cdot\mathbf{p})A(\tilde{p}_n^2)
+i\gamma_4\tilde{\omega}_n C(\tilde{p}_n^2)+B(\tilde{p}_n^2)~ .
\label{qpropt}
\ee
The quark dressing functions 
$A(\tilde{p}_n^2),B(\tilde{p}_n^2)$ and $C(\tilde{p}_n^2)$ encode effects of the
quark selfenergy $\Sigma=S^{-1}-S_0^{-1}$, expressing all deviations 
from the free propagator $S_0^{-1}$ due to nonperturbative interaction effects.
At finite temperature and chemical potential the ``shifted''
fermionic Matsubara frequencies 
$\tilde{\omega}_n = \omega_n-i\mu = (2n+1)\pi T-i\mu$ 
with temperature $T$ and chemical potential
$\mu$,  are introduced, so that 
$\tilde{p}_n = (\mathbf{p},\tilde{\omega}_n)$.
For the free quark propagator $S_0$ we have $A=C=1$ and $B=m$, the current 
quark mass.
The $\mathrm{Tr}$ operation implies summation over internal degrees of 
freedom and $\tilde{p}_n$. 
The leading term in the 2PI loop expansion is the usual one-loop
contribution which we denote as the kinetic contribution to the
thermodynamic potential
of the system 
\be
\Omega_\mathrm{kin} = \mathrm{Tr}\mathrm{Log}(S)~,
\ee
and the functional $\Psi[S]$ contains all higher loop diagrams. 
In the widely used rainbow-ladder approximation 
(see e.g. \cite{Roberts:2000aa}) 
it is given by
\be
\Psi[S]=-\frac{1}{2}\mathrm{Tr}[\Sigma S]~.
\label{rla}
\ee

A quite general and mathematically simple realization of the analytic 
structure of the quark propagator exhibits a series of CCMPs in the $p^2$ 
plane for $p^2 = -m_k^2$ and, $p^2=-m_k^{* 2}$. 
The complex numbers $m_k$ are ordered such that $|m_{k+1}/m_k|>1$.
Let us for definiteness assume that all the poles are simple. 
The following arguments can easily be generalized if the quark
 propagator has higher order poles, or branch cuts.
At zero three-momentum $\mathbf{p} = 0$ the pole structure in the 
$p^0 \equiv ip_4$ plane is a series of quartets of poles located at 
$\pm m_k = \pm m_k^R \pm i m_k^I$ 
and 
$\pm m_k^* = \pm m_k^R \mp i m_k^I$, 
and we define $m_k$ such that $m_k^R, m_k^I >0$ for all $k$.
With $\mathbf{p}\neq 0$ the poles
are given by \cite{General:2000zx}
\be
\mathcal{E}_k^2 = \mathbf{p}^2 + m_k^2 ~.
\label{energy}
\ee
where for each $k$ their locations form a quartet in the complex
energy plane at $\pm \mathcal{E}_k = \pm\epsilon_k\pm i\gamma_k$
and $\pm\mathcal{E}_k^*=\pm\epsilon_k\mp i\gamma_k$, with
\be
\begin{split}
&\epsilon_k = \frac{1}{\sqrt{2}}
\left\{(m^R_k)^2-(m^I_k)^2+\mathbf{p}^2+
\sqrt{\left[(m^R_k)^2-(m^I_k)^2+\mathbf{p}^2\right]^2
+4(m^R_k)^2(m^I_k)^2}\right\}^{1/2}\\
& \gamma_k = \frac{m^R_k m^I_k}{\epsilon_k} ~.
\end{split}  
\label{ccreim}
\ee
The analytic structure of the quark propagator governs the 
thermodynamical
properties of the system. Here we perform a simple calculation of the 
kinetic contribution to the
thermodynamic potential in the quark sector, with the proposed form
of the quark propagator.  

\subsection{Consequences for the quark sector at finite temperature}

Performing the trace in Dirac, color, flavor and momentum space, the kinetic 
term can be written as
\be
\Omega_\mathrm{kin}(T,\mu) = 
-2N_c N_f T \sum_{n=-\infty}^{+\infty}\int\frac{d^3 p}{(2\pi)^3}
\log\left[\mathbf{p}^2A^2(\tilde{p}_n^2)+
\tilde{\omega}_n^2 C^2(\tilde{p}_n^2)+B^2(\tilde{p}_n^2)\right]~ .
\label{kin1}
\ee
For simplicity we work with $N_f$ equal flavors. 
$N_c=3$ is the number of colors.

In order to perform the Matsubara sum we introduce generalized occupation 
numbers 
\be
 n_{\pm}(z)=(1+e^{\beta (z\mp\mu)})^{-1} 
\ee
having simple poles at  $z=i\tilde{\omega}_n$.
With the help of the residue theorem, the Matsubara sum is converted
to an integral along straight lines $\mathrm{Re}(z)=\mu-\delta$
and $\mathrm{Re}(z)=\mu+\delta$, where $\delta>0$ is infinitesimal 
\be
\begin{split}
I_1+I_2 &=
\int_{-i\infty+\mu+\delta}^{+i\infty+\mu+\delta} dz\, n_+(z)\log\mathcal{D}(z)+
\int_{+i\infty+\mu-\delta}^{-i\infty+\mu-\delta} dz\, n_+(z)\log\mathcal{D}(z)\\ 
&= 2\pi i \sum_{n=-\infty}^{+\infty}(-T)\log[\mathcal{D}(i\tilde{\omega}_n)]~ ,
\end{split}
\label{resid1}
\ee
where we defined
\be
\mathcal{D}(z) = \mathbf{p}^2A^2(\mathbf{p}^2,-z^2)
-z^2 C^2(\mathbf{p}^2,-z^2)+B^2(\mathbf{p}^2,-z^2)~.
\label{den}
\ee
Here and in the following we suppress the $\mathbf{p}^2$ dependence in  
$\mathcal{D}$ for brevity.

Due to the known analytic structure of the quark propagator the integrals in 
(\ref{resid1}) can be evaluated. 
We now close the contour running from 
$-i\infty+\mu+\delta$ to $+i\infty+\mu+\delta$ by
a large semicircle on the positive real axis, 
and denote this as $C_1$. Then we can rewrite
\be
\begin{split}
I_1&= T \oint_{C_1} dz 
\log\left[1+e^{-\beta (z-\mu)}\right]\frac{\mathcal{D}'(z)}{\mathcal{D}(z)}\\
&= T (-2\pi i) \sum_{k,\epsilon_k>\mu} 
\left\{\log\left[1+e^{-\beta (\mathcal{E}_k-\mu)}\right]
+\log\left[1+e^{-\beta (\mathcal{E}_k^*-\mu)}\right]\right\} ~,
\end{split}
\ee
where the first equality follows from partial integration. 
The last line is the result of the residue theorem,
and $\mathcal{E}_k$ are the previously defined poles.

For the second term we first make use of the clockwise oriented contour 
$C_2$ defined as a rectangle having vertices in 
$(+i\infty+\mu-\delta,-i\infty+\mu-\delta,-i\infty,+i\infty)$ to obtain
\be
I_2= T \oint_{C_2} dz 
\log\left[1+e^{-\beta (z-\mu)}\right]\frac{\mathcal{D}'(z)}{\mathcal{D}(z)}-
\int_{-i\infty}^{+i\infty}dz\, n_+(z)\log\mathcal{D}(z)~ .
\label{9}
\ee
If the second term in (\ref{9}) is rewritten using $n_+(z) = 1-n_-(-z)$, 
the first of the two resulting terms can be Wick rotated to 
the real axis, providing the vacuum contribution, while the second term can be 
evaluated by yet another contour, defined as $C_3$, where we close the line
running from $-i\infty$ to $+i\infty$ by a large semicircle on the negative 
real axis. 
This gives
\be
\begin{split}
I_2&= T \oint_{C_2} dz 
\log\left[1+e^{-\beta (z-\mu)}\right]\frac{\mathcal{D}'(z)}{\mathcal{D}(z)}
+T \oint_{C_3} dz 
\log\left[1+e^{-\beta (-z+\mu)}\right]\frac{\mathcal{D}'(z)}{\mathcal{D}(z)}\\
&-i\int_{-\infty}^{+\infty}dp_4\log\mathcal{D}(ip_4)=\\
&=T (-2\pi i) \sum_{k,0<\epsilon_k<\mu} 
\left\{\log\left[1+e^{-\beta (\mathcal{E}_k-\mu)}\right]
+\log\left[1+e^{-\beta (\mathcal{E}_k^*-\mu)}\right]\right\}\\
&+T (-2\pi i) \sum_{k,\epsilon_k>0} 
\left\{\log\left[1+e^{-\beta (\mathcal{E}_k+\mu)}\right]
+\log\left[1+e^{-\beta (\mathcal{E}_k^*+\mu)}\right]\right\}
-i\int_{-\infty}^{+\infty}dp_4\log\mathcal{D}(ip_4)~.
\end{split}
\ee
Collecting the obtained formulas, we can state the kinetic contribution to the 
pressure
\be
\begin{split}
\Omega_\mathrm{kin}(T,\mu) = \Omega_{\mathrm{zpt}}
- 2 T N_c N_f \sum_{k=1}^{\infty} \int \frac{d^3 p}{(2\pi)^3} 
\Big\{&\log\left[1+e^{-\beta (\mathcal{E}_k-\mu)}\right] +
 \log\left[1+e^{-\beta(\mathcal{E}^*_k-\mu)}\right]\\
 +&\log\left[1+e^{-\beta (\mathcal{E}_k+\mu)}\right] +
 \log\left[1+e^{-\beta(\mathcal{E}^*_k+\mu)}\right]\Big\} \, ,
\label{pkin}
\end{split}
\ee 
where $\Omega_\mathrm{zpt}$ is the (divergent) zero-point energy contribution
\be
\Omega_\mathrm{zpt}=
-2N_c N_f \int\frac{d^4 p}{(2\pi)^4}\log\left[p^2 A^2(p^2)+B^2(p^2)\right]~ .
\ee
It is plain to see that in the special case of just one pair of real poles at
$\pm m$, the original dispersion (\ref{ccreim}) is reduced to the one
of a free relativistic particle, and, accordingly,
the second term in Eq. (\ref{pkin}) is reduced to the free Fermi gas 
expression, a situation also encountered, e.g., in the Nambu Jona-Lasinio 
(NJL) model \cite{Nambu:1961tp,Nambu:1961fr}, 
see, e.g., \cite{Klimt:1989pm,Klevansky:1992qe,Hatsuda:1994pi,Buballa:2003qv}.
In the first term, the integral over $p_4$ can be evaluated as well, leading to
$$\Omega_\mathrm{zpt} = -4N_c N_f \sum_{k=1}^{\infty}\int\frac{d^3 p}{(2\pi)^3}
\left(\frac{\mathcal{E}}{2}+\frac{\mathcal{E}^*}{2}\right) ~,$$
which, again in the case of a pair of real poles, is just the usual zero-point
energy.

By combining the logarithms, (\ref{pkin}) can be cast in a more transparent form
\be
\begin{split}
\Omega_{\mathrm{kin}}(T,\mu) = \Omega_\mathrm{zpt}
 - 2 T N_c N_f \sum_{k=1}^{\infty} \int \frac{d^3 p}{(2\pi)^3} 
\Big\{&\log\left[1+2 e^{-\beta (\epsilon_k-\mu)}\cos(\beta\gamma_k)
+e^{-2\beta (\epsilon_k-\mu)}\right]\\
+&\log\left[1+2 e^{-\beta (\epsilon_k+\mu)}\cos(\beta\gamma_k)
+e^{-2\beta (\epsilon_k+\mu)}\right]\Big\}~ .
\end{split}
\label{pkin2}
\ee
Note that the oscillating cosine functions in the 
thermodynamic potential could render the quark matter unstable. 
Their origin is traced back to the appearance of imaginary parts 
$\gamma_k$ of the quark mass poles.
 
\subsection{Introducing the Polyakov loop}

The traced PL $\Phi(\mathbf{x},T)$ and its conjugate $\bar{\Phi}(\mathbf{x},T)$ 
are order parameters for 
confinement in quenched QCD \cite{Polyakov:1978vu,Meisinger:1995ih}, and 
as such represent important configurations of the gluon field
that should be accounted for in the effective thermodynamic description of QCD.
They are given as thermal expectation values
\be
\Phi = \frac{1}{N_c}\langle\mathrm{tr}_c(\mathcal{P})\rangle_\beta \, , \quad
\bar{\Phi} = \frac{1}{N_c}\langle\mathrm{tr}_c(\mathcal{P}^\dag)\rangle_\beta~,
\label{ploop}
\ee
where $\mathcal{P}$ is the untraced PL.
In the Polyakov gauge \cite{Polyakov:1978vu} the latter takes a simple form 
$\mathcal{P}=e^{i(\lambda_3\phi_3+\lambda_8\phi_8)}$,
where $\lambda_{3,8}$ are color Gell-Mann matrices, with $\phi_{3,8}$
being the background gluon field.
Quark (antiquark) confinement is then signalled
by $\Phi=0$ ($\bar{\Phi}=0$). 

Coupling of the PL to the quarks amounts to a modification of the quark 
occupation number function 
\be
n_\pm(z)\to\left\{1+e^{\beta\left[z\mp(\mu-i(\lambda_3\phi_3
+\lambda_8\phi_8))\right]}\right\}^{-1} ~.
\label{qnum}
\ee 
Following the same steps as in the previous subsection, the kinetic 
contribution to the thermodynamic potential can be written as 
\be
\begin{split}
\Omega_{\mathrm{kin}}(T,\mu) = 
-2 N_f T\sum_{k=1}^{\infty} \int \frac{d^3 p}{(2\pi)^3} 
\mathrm{tr}_\mathrm{c}\Big\{
&\log\left[1+\mathcal{P}e^{-\beta (\mathcal{E}_k-\mu)}\right]+
\log\left[1+\mathcal{P}e^{-\beta (\mathcal{E}_k^*-\mu)}\right]\\
+&\log\left[1+\mathcal{P}^\dag e^{-\beta (\mathcal{E}_k+\mu)}\right]
+\log\left[1+\mathcal{P}^\dag e^{-\beta (\mathcal{E}_k^*+\mu)}\right]\Big\}~ .
\label{pkinpl}
\end{split}
\ee 
Working out the color trace gives
\be
\begin{split}
\Omega_{\mathrm{kin}}(T,\mu) = 
&-2 N_f T\sum_{k=1}^{\infty} \int \frac{d^3 p}{(2\pi)^3} 
\Big\{\log\left[1+3\Phi e^{-\beta (\mathcal{E}_k-\mu)}
+3\bar{\Phi} e^{-2\beta (\mathcal{E}_k-\mu)}+
e^{-3\beta (\mathcal{E}_k-\mu)}\right]\\
&+\log\left[1+3\Phi e^{-\beta (\mathcal{E}_k^*-\mu)}
+3\bar{\Phi} e^{-2\beta (\mathcal{E}_k^*-\mu)}+
e^{-3\beta (\mathcal{E}_k^*-\mu)}\right]+(\mu\to -\mu)\Big\}~ .
\end{split}
\label{pkinpl3}
\ee 
Again, in the special case of just one pair of real poles at
$\pm m$, this expression is reduced to the corresponding term of the 
Polyakov-Nambu-Jona-Lasinio (PNJL) model
(see, e.g., \cite{Fukushima:2003fw,Ratti:2005jh,Roessner:2006xn,Fukushima:2008wg}).

The logarithms can be combined to obtain
\be
\begin{split}
\Omega_{\mathrm{kin}}(T,\mu) = 
&-2 N_f T \sum_{k=1}^{\infty} \int \frac{d^3 p}{(2\pi)^3} 
\Bigg\{\log \Big[1+6\Phi\left(e^{-\beta(\epsilon_k-\mu)}\cos(\beta\gamma_k)
+e^{-4\beta(\epsilon_k-\mu)}\cos(2\beta\gamma_k)\right)\\
+&6\bar{\Phi}\left(e^{-2\beta(\epsilon_k-\mu)}\cos(2\beta\gamma_k)
+e^{-5\beta(\epsilon_k-\mu)}\cos(\beta\gamma_k)\right)
+9\Phi^2e^{-2\beta(\epsilon_k-\mu)}+9\bar{\Phi}^2e^{-4\beta(\epsilon_k-\mu)}\\
+&18\Phi\bar{\Phi}e^{-2\beta(\epsilon_k-\mu)}\cos(\beta\gamma_k)+
2e^{-3\beta(\epsilon_k-\mu)}\cos(3\beta\gamma_k)+e^{-6\beta(\epsilon_k-\mu)}
\Big] +(\mu\to -\mu)\Bigg\}~.
\end{split}
\label{pkinpl2}
\ee
Comparing this with Eq.~(\ref{pkin2}), we see that now the dominant cosine 
terms are weighted by the PL. As a consequence, the pressure instabilities
are highly suppressed in the confined phase: As long as $\Phi$ and $\bar\Phi$
are zero, there remains only one cosine term, which is, however, suppressed  
by the third power of the Boltzmann factor. 
In fact, the mechanism is basically the same as in the PNJL model, where
the coupling to the PL suppresses the quark degrees of freedom
at low $T$, but does not eliminate them entirely \cite{Ratti:2005jh,
Roessner:2006xn,Fukushima:2008wg}. 

\subsection{Zero temperature, finite chemical potential}

In this part, special attention is devoted to the effects of
the CCMPs along the $T=0$, $\mu>0$ axis. 
The Matsubara sum in (\ref{kin1}) gets converted to an integral 
over $p_4$. 
\be
\Omega_\mathrm{kin}(0,\mu) = -2N_c N_f\int\frac{d^4 p}{(2\pi)^4}
\log\left[\mathbf{p}^2 A^2(\tilde{p}^2)+\tilde{p}_4^2
C^2(\tilde{p}^2)+B^2(\tilde{p}^2)\right]\,~
\label{kinmu}
\ee
where
$$\tilde{p}^2 = \mathbf{p}^2+\tilde{p}_4^2 , 
\quad  \tilde{p}_4=p_4-i\mu ~ .$$
We start by considering the quark number density
\be
n(\mu)=-\frac{\partial\Omega_\mathrm{kin}}{\partial\mu} 
= 2 N_f N_c\int\frac{d^4 p}{(2\pi)^4}
(-2i\tilde{p}_4)
\frac{\partial\mathcal{D}(i\tilde{p}_4)}{\partial\tilde{p}_4^2}
\frac{1}{\mathcal{D}(i\tilde{p}_4)}
~,
\label{dens1}
\ee
where $\mathcal{D}(i\tilde{p}_4)$ is given by (\ref{den}).
At zero chemical potential, the quark density is zero,
as it is obvious from 
the integrand being an odd function of $p_4$. 
This allows for the evaluation of the integral by a clockwise oriented 
rectangular contour in the complex $p_4$ plane having vertices in 
$(-\infty,\infty,\infty-i\mu,-\infty-i\mu)$.
As the poles (\ref{ccreim}) are defined in Minkowski space,
in Euclidean space this means that the only poles that
enter the contour have 
$\mathrm{Re}(\mathcal{E}_k)<\mu$. 
Therefore
\be
\oint_C \frac{dp_4}{2\pi}
(-2ip_4)
\frac{\partial\mathcal{D}(ip_4)}{\partial p_4^2}
\frac{1}{\mathcal{D}(ip_4)}
=-2\pi i\sum_{k=1}^\infty\left[\mathrm{Res}(-i\mathcal{E}_k)
+\mathrm{Res}(-i\mathcal{E}_k^*)\right]\theta(\mu-\epsilon_k)~ .
\label{cont}
\ee
The residue can straightforwardly be shown to be $1/2\pi i$ in both cases, 
giving
\be
n(\mu)=4 N_f N_c\sum_{k=1}^{\infty}\int\frac{d^3 p}{(2\pi)^3}
\theta(\mu-\epsilon_k)~.
\ee
The theta function defines a ``generalized'' Fermi momentum
\be
p_F(\mu,m_k^R,m_k^I) = 
\mu\sqrt{\left[1-\frac{(m_k^R)^2}{\mu^2}\right]
\left[1+\frac{(m_k^I)^2}{\mu^2}\right]}~.
\label{ferm}
\ee
The quark number density can now be obtained as
\be
n(\mu) = \frac{2N_f N_c}{3\pi^2}\sum_{k=1}^\infty
p_F^3(\mu,m_k^R,m_k^I)\theta(\mu-m_k^R)~.
\label{dens2}
\ee
It is actually remarkable that the density thresholds depend
only on the real parts, $m_k^R$.
The imaginary parts $m_k^I$ enhance the Fermi momenta 
(and thus the density) compared to the values one would get for $m_k^I=0$. 
For $m_k^I>m_k^R$ and 
$$\frac{1}{\mu^2}<\frac{1}{(m_k^R)^2}-\frac{1}{(m_k^I)^2}$$
the Fermi momentum is even larger than $\mu$. 
This point will later become important.

Integrating the expression
$$\Omega_\mathrm{kin}(0,\mu) = -\int^\mu d\mu' n(\mu') 
= -\frac{N_f N_c}{3\pi^2}\sum_k\int_{m_k^R}^\mu d\mu' 
p_F^3(\mu',m_k^R,m_k^I)$$
the thermodynamic potential can be reconstructed in a closed form
$$\Omega_\mathrm{kin}(0,\mu) = -\frac{2N_f N_c}{3\pi^2}\sum_{k=1}^{\infty} 
\omega(\mu,m_k^R,m_k^I)~ ,$$
where
\be
\begin{split}
\omega(\mu,x,y)=&-\frac{p_F}{8\mu}[4x^2 y^2+5(y^2-x^2)\mu^2+2\mu^4]+
\frac{3}{16}(x^4-6x^2y^2+y^4)
\log\left[\frac{y^2-x^2+2\mu(p_F+\mu)}{x^2+y^2}\right]\\
&-\frac{3}{4}xy(y^2-x^2)\arctan
\left[\frac{2xy\mu p_F}{(x^2-y^2)\mu^2+2x^2 y^2}\right]~ .
\end{split}
\label{smom}
\ee
It is straightforward to see that in the case of only a pair of real mass 
poles $\pm m$ we get the familiar expression for the free, massive,
relativistic Fermi gas
\be
\Omega_\mathrm{kin}(0,\mu) = 
-\frac{N_f N_c}{3\pi^2}\frac{1}{8} \left[2\mu^3 p_F -5m^2\mu p_F 
+3m^4\log\left(\frac{p_F+\mu}{m}\right)\right] \, .
\label{sim}
\ee
We mention that at zero temperature the PL decouples, so it has no effect on the EoS.

\section{Instabilities in a non-local chiral quark model}

The dispersion relations $\mathcal{E}_k$ which enter
Eq. (\ref{pkin}) are governed by the analytic structure
of the quark propagator, so that further insight 
can be obtained only by studying the 
thermal behavior of the quark propagator, i.e., by understanding how
the CCMPs respond to a change in the temperature or density.

Parametrizing the analytic structure, say, from lattice studies
at finite $T$ is very demanding.
In this case the analytic structure is also somewhat arbitrary as 
the quark propagator is known only at a finite number of points, allowing 
for different meromorphic forms \cite{Alkofer:2003jj}.
For the present purpose we will therefore study a specific model as an
example case. 
More precisely, we consider a Dyson-Schwinger model with a \textit{separable} 
gluon interaction 
\cite{Plant:1997jr,Blaschke:1999ab,Blaschke:2000gd}.
In the rainbow-ladder approximation, these models
are in fact identical to mean-field non-local NJL models,
see, e.g., \cite{Hell:2008cc,Contrera:2010kz,Horvatic:2010md}.
They capture the important aspect of momentum dependent
dressing functions in the quark propagator (\ref{qpropt}) by introducing 
regulator functions which also ensure the convergence of loop integrals.
Here we consider the particularly simple rank-1 case, where $A=C=1$, while
\be
B(p^2) = m+bf_0(p^2)~,
\label{mrun}
\ee
with $b$ being the chiral symmetry breaking parameter (mass gap) and
$f_0(p^2)$ the regulator function.
The latter is an input of the model.

For this kind of separable models, 
it was already observed in the literature that pressure instabilities 
appear in certain regions of the $T$-$\mu$ plane.
In Refs.~\cite{Blaschke:1999ab,General:2000zx,GomezDumm:2001fz,
GomezDumm:2004sr,Loewe:2011qc} this was found in full numerical studies, 
and in \cite{Loewe:2011qc}
also by restricting the calculations to a finite number of CCMPs. 
The aim of the present section is to demonstrate that these
instabilities are driven by the presence of the CCMPs, and then
to study the effect of the PL.  
Results from a full numerical study will be confronted with a calculation
where we restrict ourselves only to a finite number of CCMPs, 
demonstrating that the instability region is actually completely
dominated by the first quartet.

\subsection{Analytic structure}

\begin{figure}
\begin{center}
\psfig{file=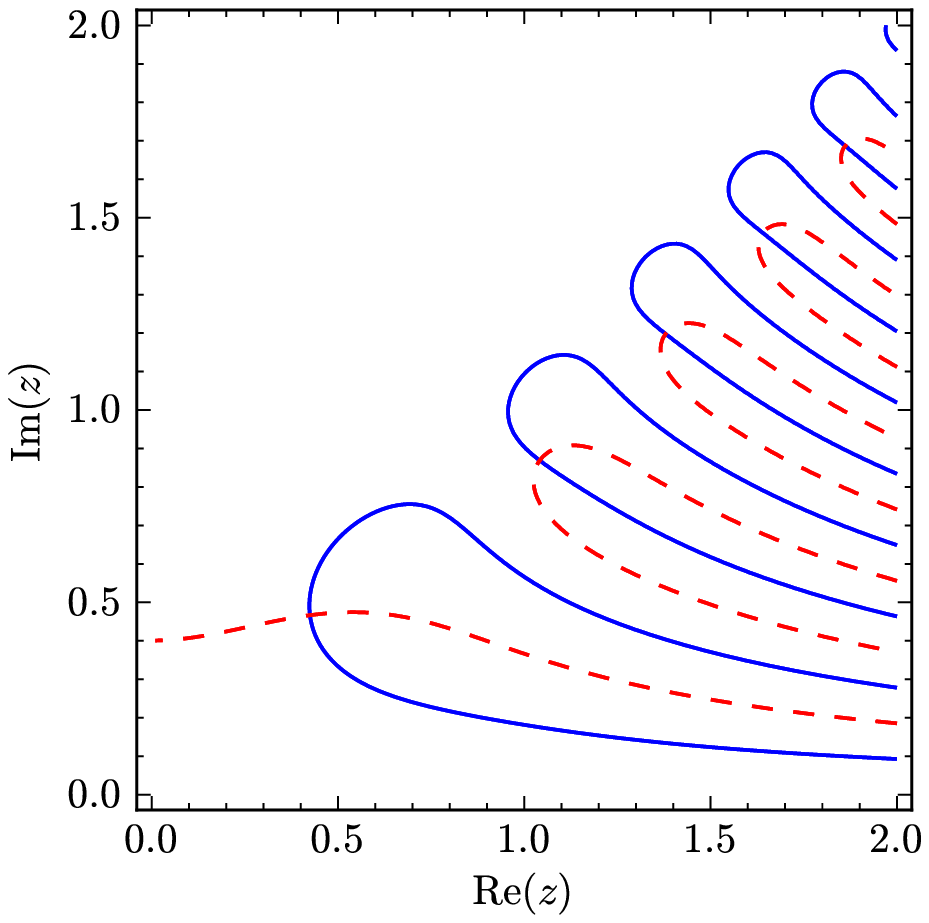,scale=0.6}
\psfig{file=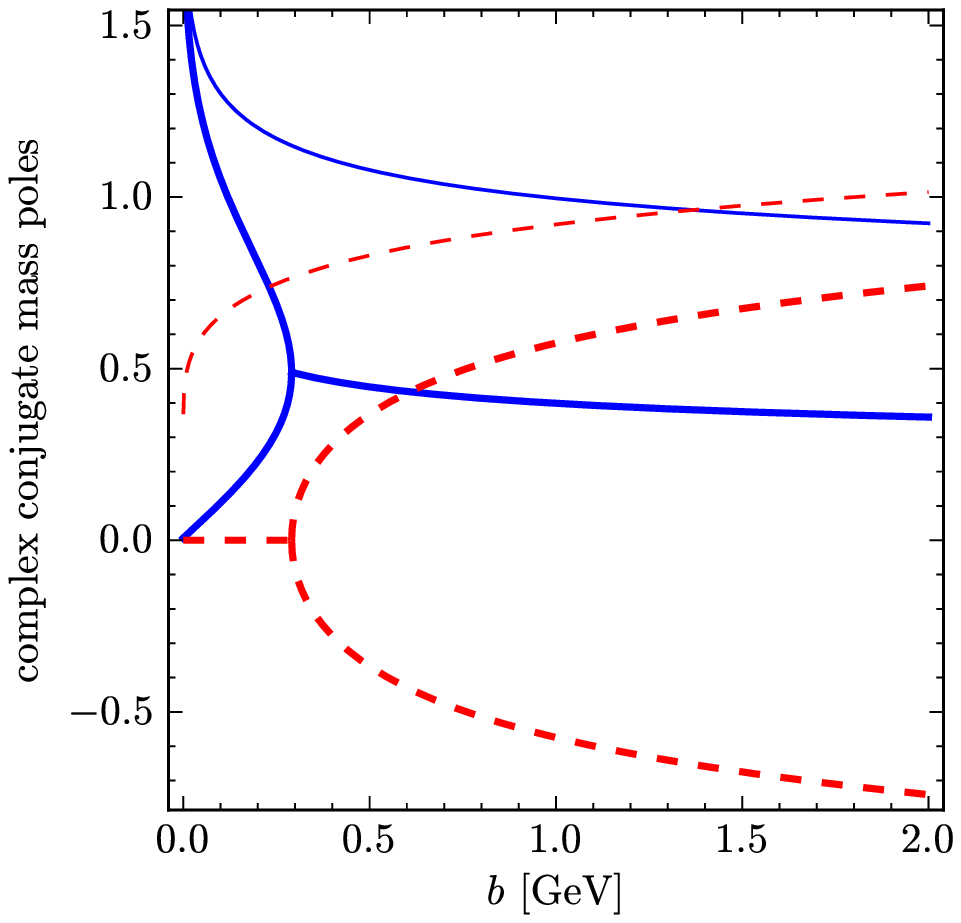,scale=0.6}
\caption{(Color online) Left: The curves where the real
(blue, solid) and imaginary (red, dashed) parts of the propagator denominator
$\mathcal{D}=-z^2+B^2(-z^2)$ vanish for 
the value of the mass gap $b$ in the 
vacuum.
Right: Real parts (blue, solid) and the corresponding imaginary 
parts (red, dashed) of the quark propagator mass poles as functions
of the gap parameter $b$.
The lowest lying poles correspond to the thick lines, 
the next higher lying quartet to the thin lines.}        
\label{ccpoles}
\end{center}
\end{figure}

The analytic structure of the model was detailed in 
Refs.~\cite{General:2000zx,GomezDumm:2001fz,GomezDumm:2004sr} for
Gaussian and Lorentzian 
regulators.
We briefly summarize their analysis for the case of the Gaussian regulator,  
given by
\be
f_0(p^2) = e^{-p^2/\Lambda_0^2}~,
\label{formf}
\ee
with a parameter $\Lambda_0$.
The quark propagator has then an infinite number of CCMPs, as exemplified 
on the left plot in Fig. \ref{ccpoles}. 
The position of the poles is controlled by the value of the gap. 
If the gap $b$ is larger than a critical value $b_c$ given by 
\be
b_c = \frac{1}{2}\left(\sqrt{m^2+2\Lambda_0^2}-m\right)
e^{-\sqrt{m^2+2\Lambda_0^2}/4\Lambda_0^2}~ ,
\label{gapcrit}
\ee
all the poles are
complex. 
For the quartet nearest to the origin, an especially interesting situation 
occurs. 
As the gap gets smaller, the poles travel to the real axis, where they meet in 
doublets at $b = b_c$. 
If the gap is further reduced, $b<b_c$, every doublet again splits, 
with one pole eventually
going to plus (minus) infinity and the other arriving at $m$ ($-m$),
for $b=0$. 
At that point, the real parts of the higher quartets go to infinity, while 
the imaginary
parts go to zero. See Fig.~\ref{ccpoles} for the behavior of the first
and the second quartet w.r.t.\@ the mass gap. 

In our numerical calculations we adopt the 
parameters of Ref.~\cite{Blaschke:1999ab}, $\Lambda_0 = 0.687$ GeV, 
$m = 0.0096\Lambda_0$, and $D_0 = 128/\Lambda_0^2$, successfully reproducing 
low-energy phenomenology. Here $D_0$ is the 
strength of the non-local effective gluon interaction 
(i.e., four-quark non-local NJL interaction).
For these parameters one obtains $b_c = 0.295$ GeV, while 
the vacuum solution of the gap equation is $b_\mathrm{vac} =0.678$ GeV.
Thus, the gap is overcritical in this case.

\subsection{Thermodynamic potential and
in-medium mass gap}

The in-medium properties of the model are obtained from the 
thermodynamic potential (\ref{CJT}), coupled to the PL.
In mean-field approximation (\ref{rla}) it is given by
\be
\Omega(b,\Phi,\bar{\Phi}) = \Omega_\mathrm{cond}(b)+
\Omega_\mathrm{kin}(b,\Phi,\bar{\Phi})+\mathcal{U}(\Phi,\bar{\Phi})~,
\label{tpotfull}
\ee
see, e.g., \cite{Blaschke:1999ab}.
Here $\Omega_\mathrm{cond}=N_f \frac{9}{8D_0}b^2$ 
and $\mathcal{U}$ represents the mean-field PL potential, for which we use
the familiar polynomial form found in Ref.~\cite{Ratti:2005jh}.
Other forms of this potential are in use, like the logarithmic form
\cite{Roessner:2006xn}, a strong-coupling inspired one \cite{Fukushima:2008wg} 
or a $\mu$-dependent one \cite{Dexheimer:2009va}.
For recent developments, see 
\cite{Sasaki:2012bi,Ruggieri:2012ny,Fukushima:2012qa}.

$\Omega_\mathrm{kin}$ is provided by Eq.~(\ref{kin1}) 
augmented with the PL. This amounts to
\be
\Omega_\mathrm{kin}(b,\Phi,\bar{\Phi}) 
= -2N_c N_f T
\sum_{n=-\infty}^{+\infty}\int\frac{d^3 p}{(2\pi)^3}
\mathrm{tr}_c\log\left[\tilde{p}_n^2
+B^2(\tilde{p}_n^2)\right]~ ,
\label{kin2}
\ee
where we now understand $\tilde{p}_n^2$ as diagonal matrices in color space
\be
\tilde{p}_n^2 = \mathbf{p}^2+\tilde{\omega}_n^2~, \quad 
\tilde{\omega}_n(\phi_3,\phi_8)=\omega_n-i\mu
+\lambda_3\phi_3+\lambda_8\phi_8~ .
\label{pcol}
\ee
To simplify the calculations, a further restriction is imposed by setting
$\phi_8=0$, i.e., $\Phi=\bar{\Phi}$.
The thermal properties of the model then follow from the minimization of 
the thermodynamic potential w.r.t.\@ $b$ and $\phi_3$. 
In particular, the $T$ and $\mu$ dependence of the dressing function $B$ is 
solely determined by the mass gap $b$, as we have seen above.
The explicit form of the mass gap equation is presented in the appendix.

\begin{figure}
\begin{center}
\parbox{15cm}{
\psfig{file=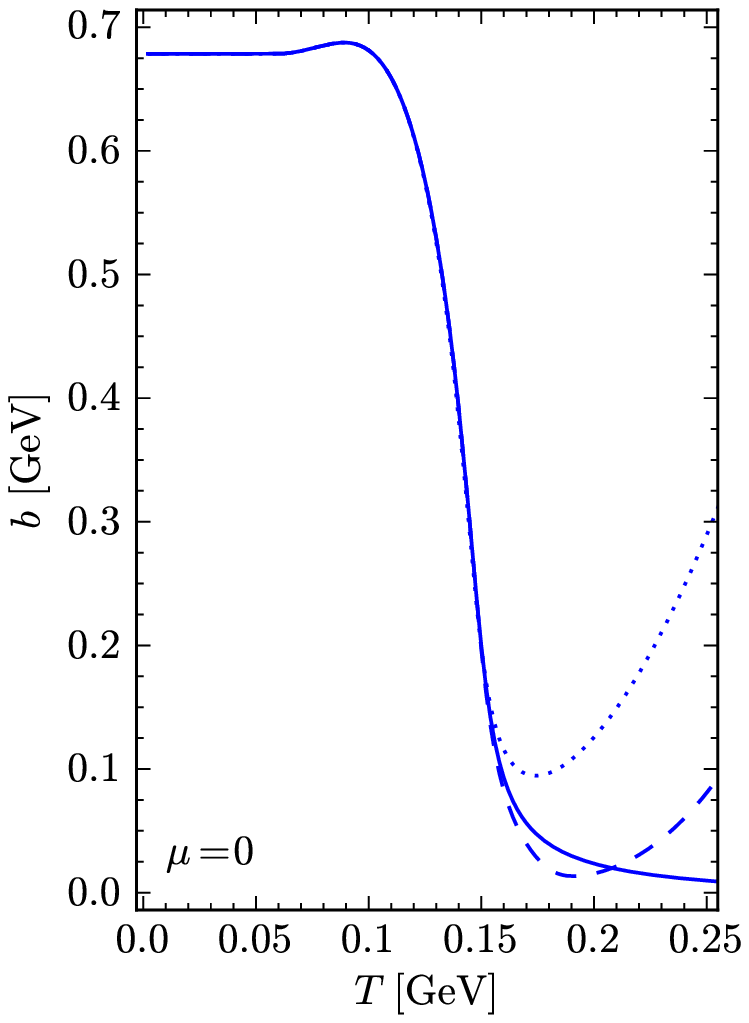,scale=0.6}
\psfig{file=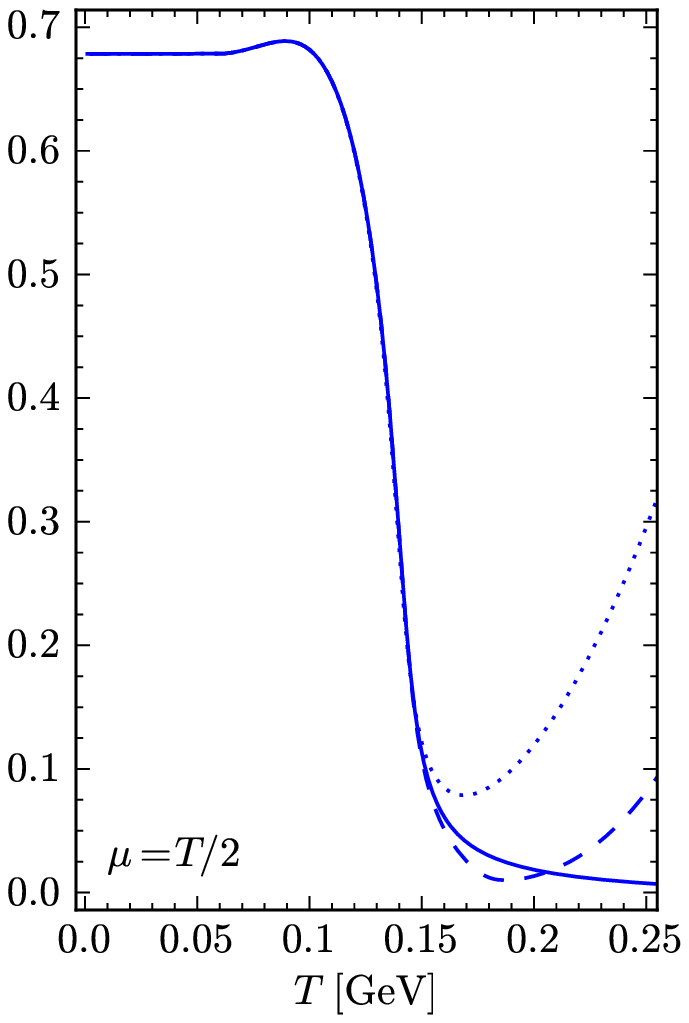,scale=0.6}
\psfig{file=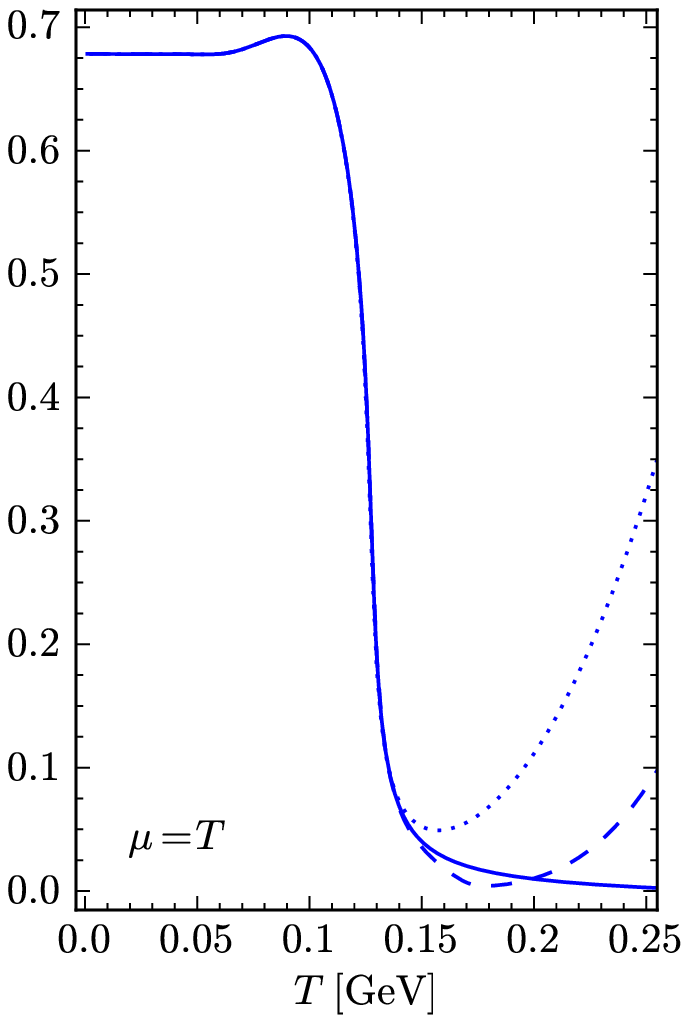,scale=0.6}
}
\caption{(Color online) The mass gap as a function
of temperature along lines of constant $\mu/T$, for a system
without the PL.
The full line is the complete numerical calculation, while the dotted
and the dashed lines correspond to the approximation 
accounting for the first and the first plus second quartet of poles, 
respectively.}        
\label{btmp}
\end{center}
\end{figure}

In Fig. \ref{btmp} the mass gap of a system without PL is shown 
as a function of the temperature along different lines of constant $\mu/T$.
The results of a full numerical solution (full lines) are compared with 
approximate ones where only a small number of poles is taken into
account. 
The dotted lines indicate calculations where the system is approximated 
by only the lowest lying quartet, containing
the states that become physical quark degrees of freedom
when $b$ drops below $b_c$.
In order to demonstrate the convergence towards the full numerical results,
we also show the effect of additionally including the second quartet 
(dashed lines).

An observation that will be crucial later on is that, at low temperatures, 
a perfect agreement with the numerical solutions is obtained
already with the first quartet. 
Deviations start only after chiral restoration, 
so that at higher temperatures higher quartets are needed
to develop the correct chiral behavior.
In fact, for any finite number of poles, the mass gap increases again after
reaching a minimum, so that the correct high-temperature limit is only 
reached if all poles are included.

\begin{figure}
\begin{center}
\parbox{14cm}{
\psfig{file=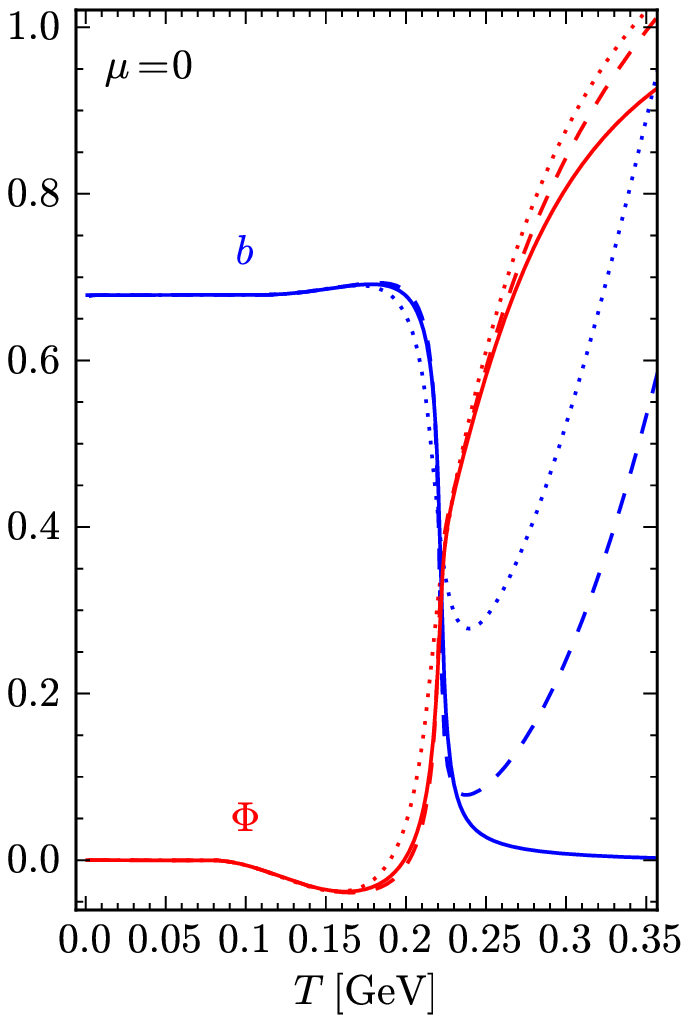,scale=0.6}
\psfig{file=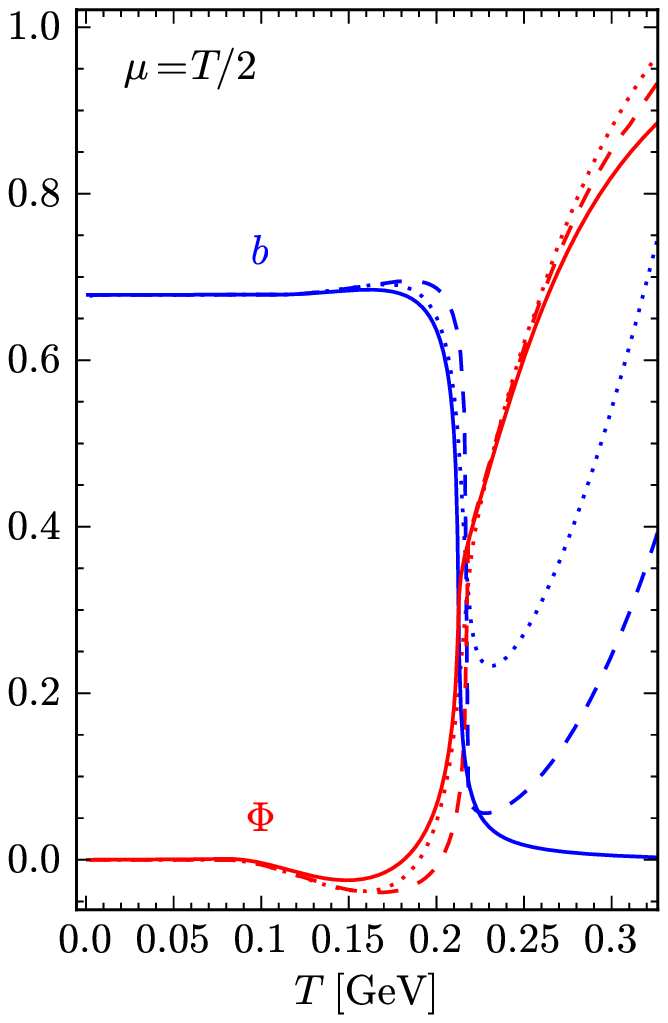,scale=0.6}
\psfig{file=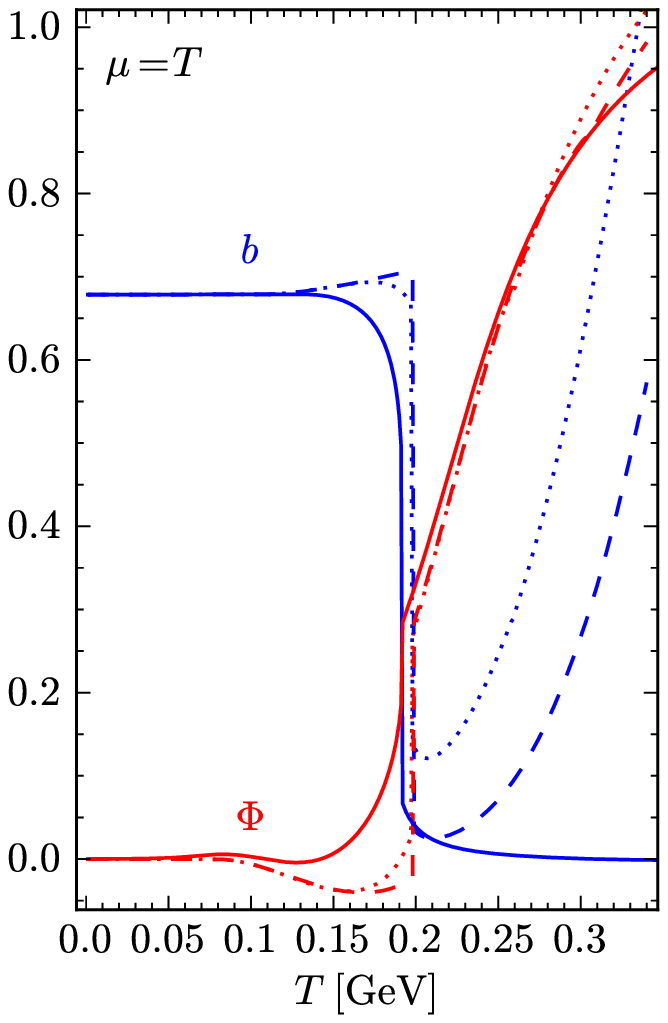,scale=0.6}
}
\caption{(Color online) 
The mass gap (blue), and the PL (red) as functions of temperature.
Line styles as in Fig.~\ref{btmp}.}        
\label{bphitmp}
\end{center}
\end{figure}

As seen in Fig.~\ref{bphitmp},
similar conclusions hold when the PL is introduced,
although the deviations are
slightly more pronounced after the chiral/deconfinement transition,
and for higher chemical potential also around the transition. 
Let us observe that at $\mu/T=1$
(rightmost plot in Fig.~\ref{bphitmp}),
the system develops a first-order transition.
 
In Fig.~\ref{bphitmp} we also show the results for the PL expectation value. 
While the overall behavior is the expected one, rising from $\Phi=0$ at low
$T$ towards $\Phi=1$ at high $T$, it turns out that there is an intermediate
regime below the cross-over temperature, where $\Phi$ gets negative. 
Although formally allowed by the definition (\ref{ploop}), which at the 
mean-field level boils down to 
$\Phi = \frac{1}{N_c} [1+2\cos\left(\beta\phi_3\right)]$,
it is in sharp contrast to the standard interpretation of $\Phi$ as 
exponential of the free energy $F_q$ of a static color source, 
$\Phi = e^{-\beta F_q}$~\cite{McLerran:1981pb}.
We also note that, roughly in the same region, the gap parameter $b$
{\it rises} as a function of $T$, a feature, which is even more pronounced
in the calculation without PL, Fig.~\ref{btmp}.

\subsection{Pressure instabilities and instability suppression}

To calculate the EoS,
the kinetic contribution to the thermodynamic
potential is regularized by
subtracting the zero-point energy of free quarks, i.e.,
\be
\Omega_\mathrm{zpt}^\mathrm{reg} = -2N_c N_f \int\frac{d^4 p}{(2\pi)^4}
\log\left[\frac{p^2+B^2(p^2)}{p^2+m^2}\right]~,
\label{rgzpt}
\ee 
and the EoS is given by $p(T,\mu)=-\Omega(T,\mu)-\Omega_0$, 
where $\Omega_0$ is a constant chosen to achieve zero pressure in the vacuum.

\begin{figure}
\begin{center}
\parbox{15cm}{
\psfig{file=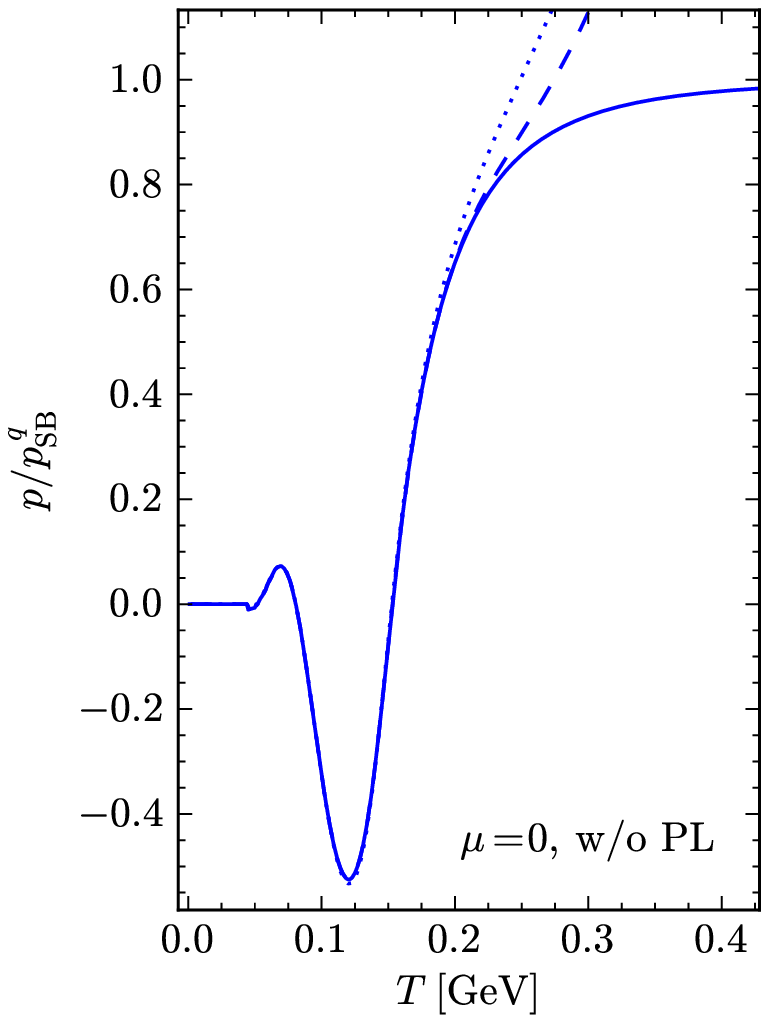,scale=0.6}
\psfig{file=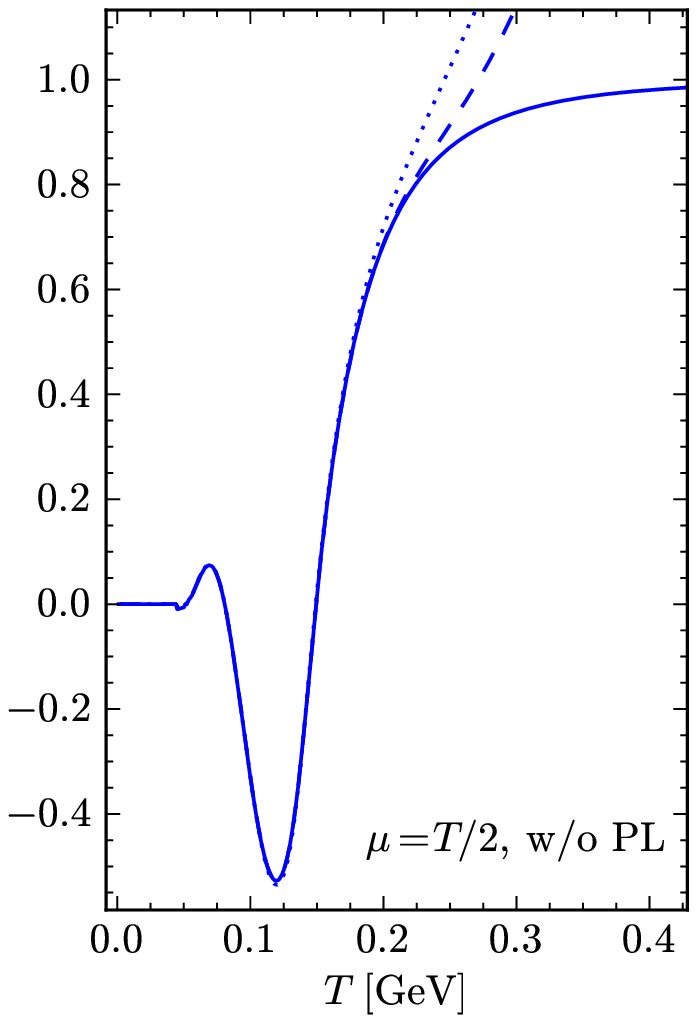,scale=0.6}
\psfig{file=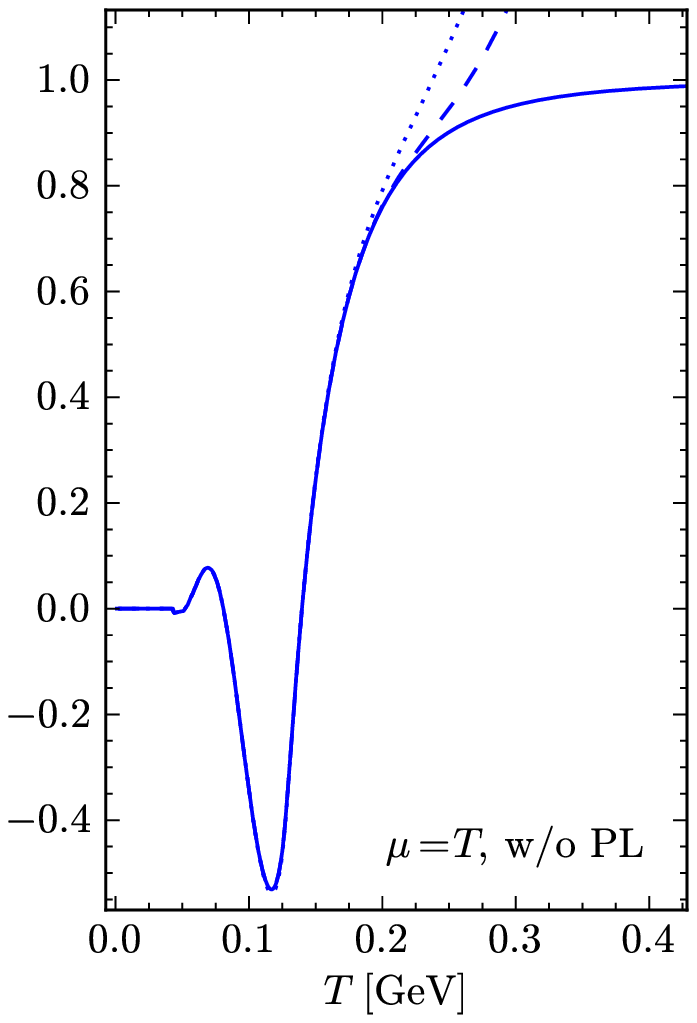,scale=0.6}
}
\caption{Scaled pressure $p/p_{SB}^q$ as a function of temperature, for a 
system without the PL. 
Line styles as in Fig.~\ref{btmp}.}     
\label{ptmp4}
\end{center}
\end{figure}

\begin{figure}
\begin{center}
\parbox{14cm}{
\psfig{file=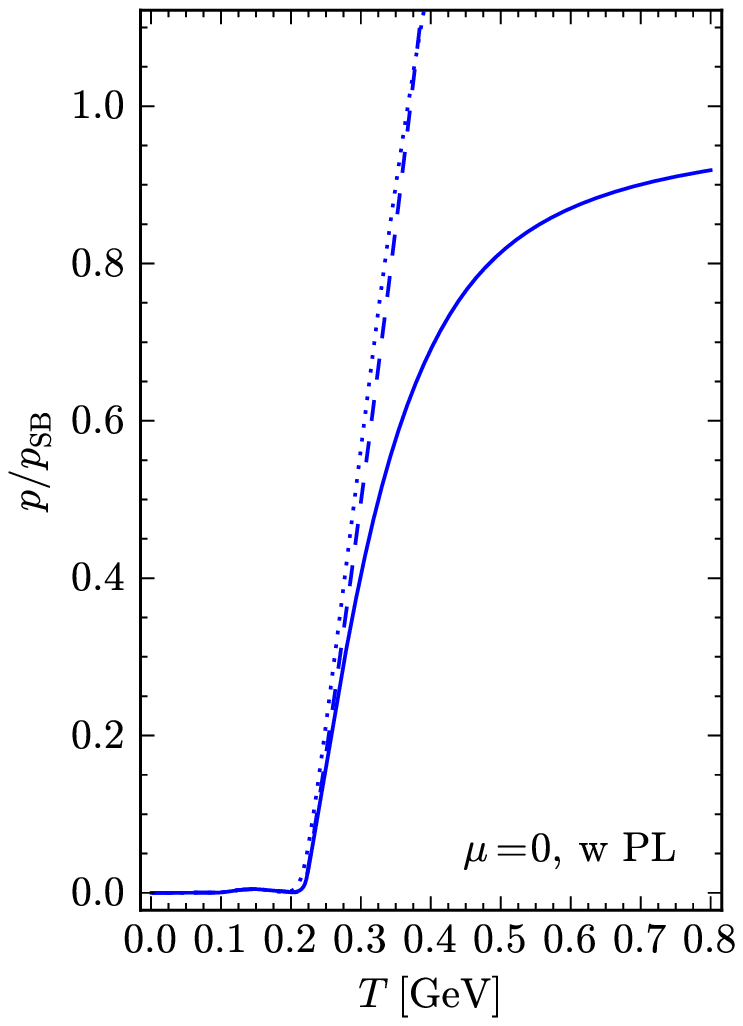,scale=0.6}
\psfig{file=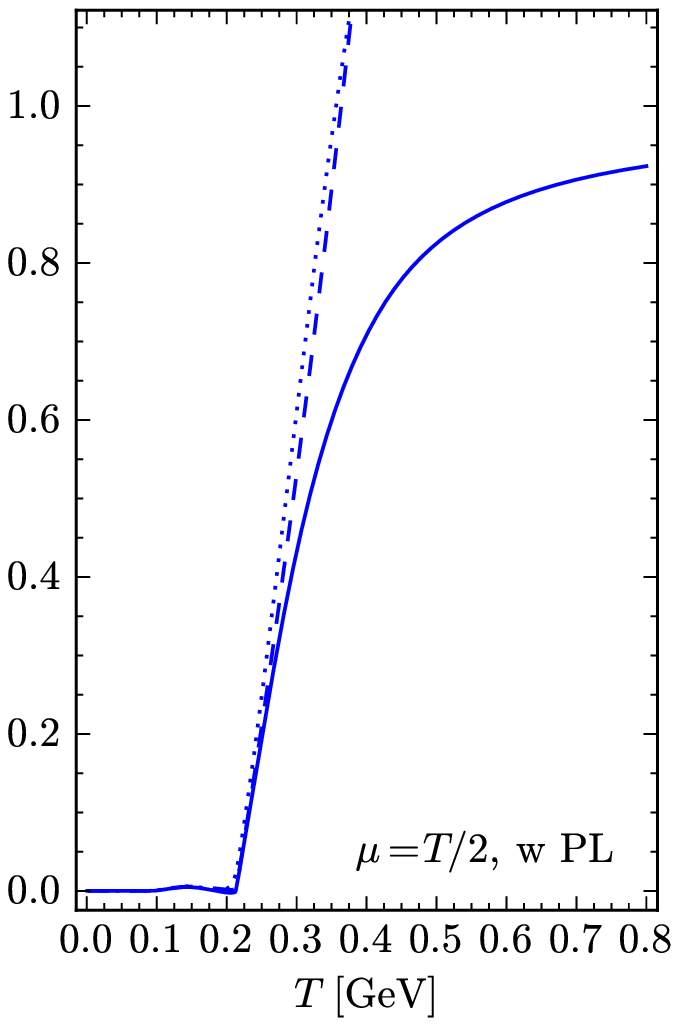,scale=0.6}
\psfig{file=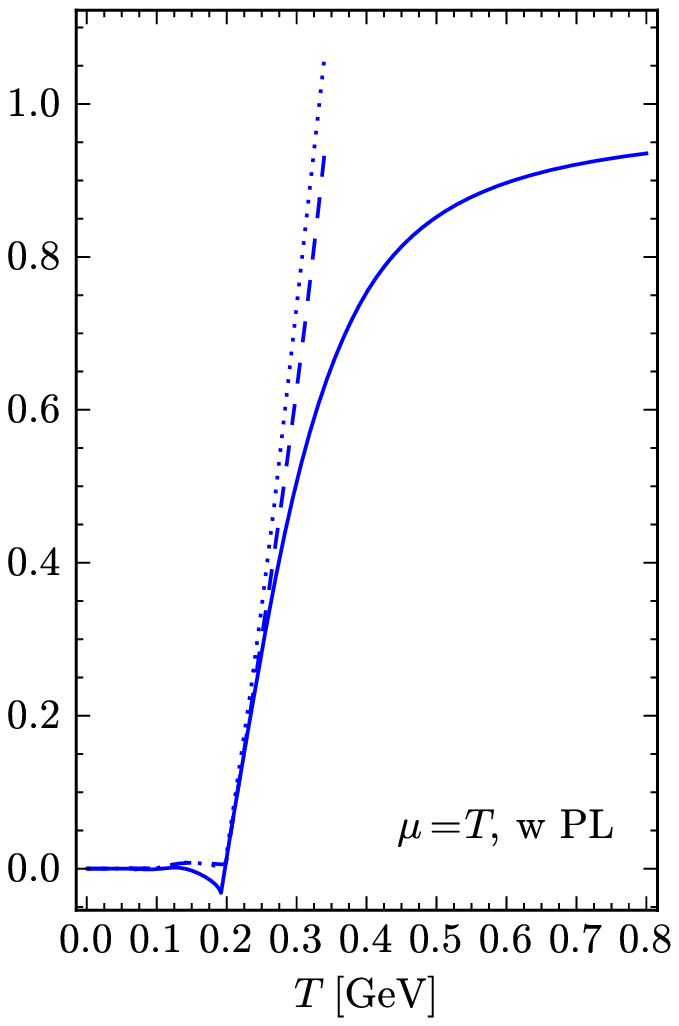,scale=0.6}
}
\caption{Scaled pressure $p/p_{SB}$ as a function of temperature, for a 
system with the PL. 
Line styles as in Fig.~\ref{btmp}.}        
\label{pphitmp4}
\end{center}
\end{figure}

In Figs.~\ref{ptmp4} and \ref{pphitmp4}, the pressure is displayed as a 
function of temperature, again along lines of constant $\mu/T$.
In the model without PL, the results are scaled by the pressure of $N_c N_f$ 
noninteracting massless quarks, 
\be
p_\mathrm{SB}^q = N_c\, N_f\left[\frac{7\pi^2}{180}+
\frac{1}{6} \, \left(\frac{\mu}{T}\right)^2
+\frac{1}{12\pi^2}\,\left(\frac{\mu}{T}\right)^4\right]T^4~,
\label{sblimq}
\ee
whereas in the case with PL, we divide by the full 
Stefan-Boltzmann (SB) pressure of
$N_c N_f$ massless quarks and $N^2_c-1$ gluons, 
\be
p_\mathrm{SB} = p_\mathrm{SB}^q + (N_c^2-1)\frac{\pi^2}{45}T^4~. 
\label{sblim}
\ee

The results for the model without PL are displayed in Fig.~\ref{ptmp4}.
The most striking features are the oscillations, which signal the
thermodynamic instabilities, we have anticipated from the 
cosine terms in Eq.~(\ref{pkin2}).
They turn out to be particularly troublesome, as there are not only 
temperature regions where the pressure drops with increasing temperature,
but where it gets even 
negative.\footnote{Let us recall that we defined the vacuum pressure 
to be zero, so this result is clearly unphysical.}
Comparing the three panels of the figure, the results seem to be rather
independent of the ratio $\mu/T$. We should keep in mind, however, 
that the pressure is scaled by the SB value, which is larger for 
larger values of $\mu/T$. Taking this into account, the instabilities 
grow with the chemical potential, 
since the Boltzmann factors are even less effective in damping the 
oscillating terms. 
This results in a rather large negative pressure for $\mu/T=1$.

For comparison we show again the results obtained when we only take into
account the lowest-lying poles. 
In agreement with our findings for the mass gap, we observe that, at low 
temperatures and more importantly, in the region of the instability, 
the pressure given by 
just the first quartet is an excellent approximation. 
The oscillations of the pressure can thus be understood quantitatively
from the temperature dependence of $b$ shown in Fig.~\ref{btmp},
together with the $b$ dependence of the lowest-lying poles shown in
the right panel of  Fig.~\ref{ccpoles}.
In particular, at $T$ around 150~MeV, the mass gap drops below $b_c$,
so that the lowest quartet splits into two real doublets and
no longer yields an oscillating behavior.  

At high temperatures, the full numerical result for the pressure 
(solid lines) approaches the SB limit, whereas the restriction to the first 
quartet (dotted) strongly overshoots this limit and is, thus, not a good 
approximation in this regime. 
The inclusion of the second quartet (dashed line) leads to some improvement 
but fails as well to reproduce the SB limit. 
This is consistent with Fig.~\ref{btmp},
where the restriction to a few mass poles even qualitatively 
failed to reproduce the high-temperature behavior of the mass gap.

Introducing the PL leads to a dramatic improvement of the EoS.
As demonstrated on Fig.~\ref{pphitmp4}, the oscillations are 
strongly suppressed.
Since the PL does not eliminate all the cosine terms completely
(see Eq.~(\ref{pkinpl2})),
residual wiggles are still present on the results for $\mu/T=0,1/2$, while
at $\mu/T=1$ also a slightly negative pressure is observed in the full numerical
calculation. 
We also note that the negative values of the PL, which we have seen in 
Fig.~\ref{bphitmp}, appear roughly in the same temperature region, but 
are more pronounced at low chemical potentials.  
So, to some extent, there seems to be a trade-off between an unphysical
behavior of the pressure and an unphysical behavior of the Polyakov loop
(when interpreted as exponential of the free energy of a static quark).

\subsection{Effect of the poles at $T=0$}

In the final part of this section, we want to discuss another consequence
of the CCMPs, which shows up in the non-local chiral model at zero temperature.
We recall that the PL decouples at $T=0$ and therefore has no effect in this
case.

\begin{figure}
\begin{center}
\parbox{14cm}{
\psfig{file=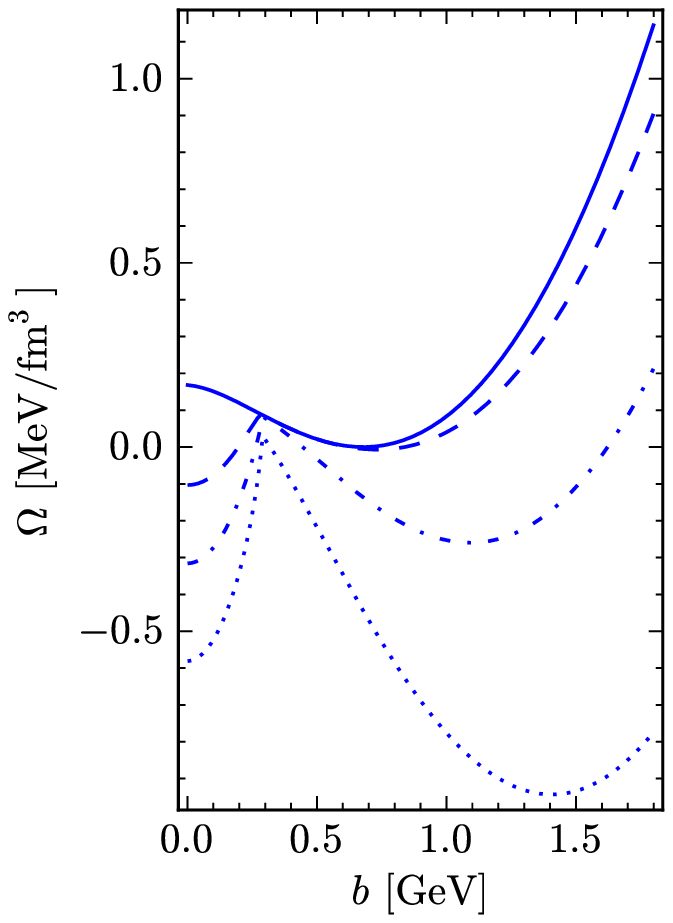,scale=0.65}
\psfig{file=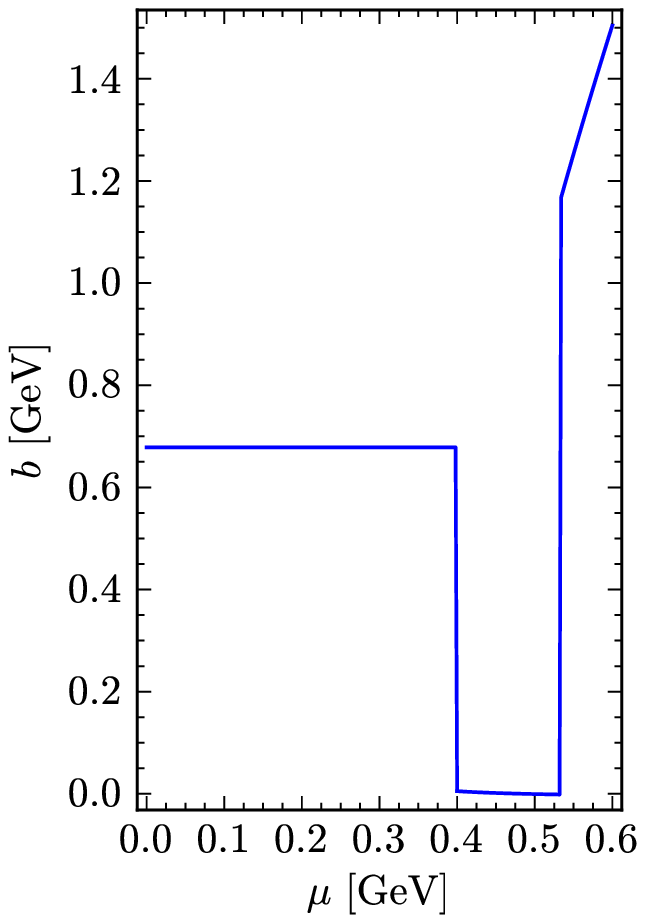,scale=0.65}
\psfig{file=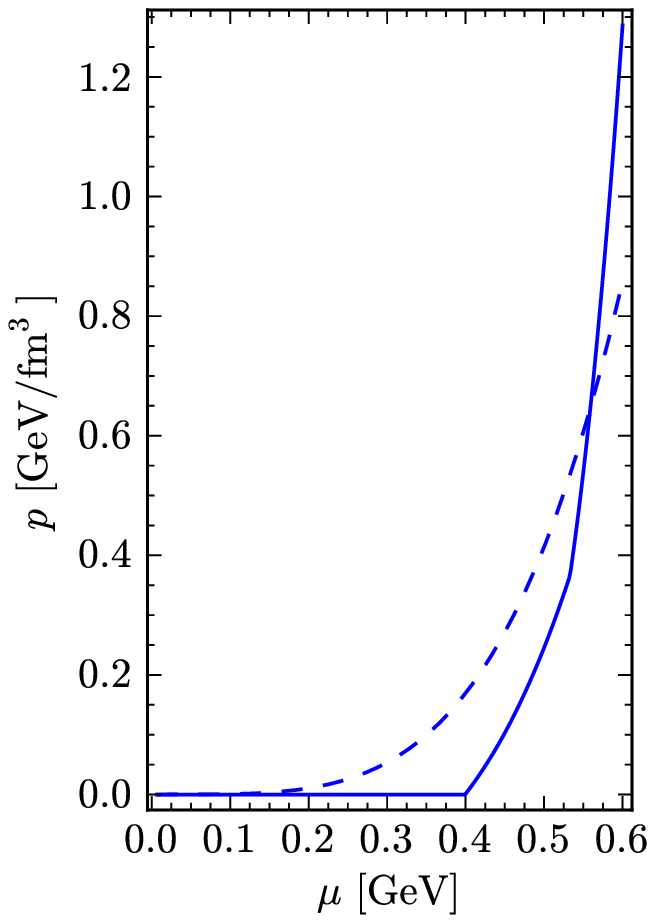,scale=0.65}
}
\caption{
(Color online) The first plot shows the behavior of the regularized
thermodynamic potential at $T=0$ as a function of the mass gap. The curves 
correspond to different chemical potentials: 
$\mu=0$ (solid), $\mu=0.45$~GeV (dashed), $\mu=0.52$~GeV (dash-dotted), 
and  $\mu=0.58$~GeV (dotted), where chiral symmetry is broken again.
This is made transparent on the second plot, where the mass gap 
is displayed as a function of the chemical potential. On the third plot
the resulting EoS is shown (full line), in comparison
with the EoS for a gas of free gas of quarks (dashed line).
}        
\label{fntmju}
\end{center}
\end{figure}

In the left panel of Fig.~\ref{fntmju}, we show the regularized thermodynamic 
potential as a function of the gap parameter $b$ for several values of the 
chemical potential $\mu$. 
According to Eq.~(\ref{dens2}), the threshold chemical potential for non-zero 
quark number density is controlled by the lowest value of $m_k^R$.
Hence, at finite chemical potential, the thermodynamic potential has to 
stay at its vacuum value in those regions, where the lowest threshold, 
$m_1^R$, is bigger than $\mu$.
This is similar to the local NJL model, where at $T=0$ the thermodynamic 
potential as a function of the constituent quark mass $M$ stays at its
vacuum value for $M > \mu$.\footnote{More general, this is a consequence
of the so-called ``Silver Blaze problem''~\cite{Cohen:2003kd}, 
meaning that at $T=0$ the grand partition function must not change
if $\mu$ is below the smallest excitation threshold.}
In the non-local model the essential difference is that the threshold 
$m_1^R$ is a non-monotonic function of the gap parameter $b$.
As we have seen in Fig.~\ref{ccpoles}, it rises from $m_1^R=m$ at $b=0$ to 
a maximum at $b=b_c$, where 
$m_1^R(b_c)= \frac{1}{2}\left(m+\sqrt{m^2+2\Lambda_0^2}\right)
= 0.489$~GeV for our parameters.
Above this point, $m_1^R$ slowly decreases, reaching arbitrarily
small values at large $b$.
As a consequence, for $0<\mu<m_1^R(b_c)$, the thermodynamic potential 
coincides with the vacuum curve only in a finite interval around $b_c$, 
bounded by the condition $m_1^R(b)=\mu$.
For the mass gap outside this range there is a finite 
density of quarks and the thermodynamic potential is below the vacuum one.
An example for this case is given by the dashed line in the left panel of 
Fig.~\ref{fntmju}.

For $\mu>m_1^R(b_c)$ the potential is reduced everywhere (dash-dotted line).
Eventually, this leads to a pathological result:
Whereas at intermediate chemical potentials a global minimum emerges near
$b=0$, leading to (approximate) chiral-symmetry restoration, at sufficiently 
high chemical potential, the nontrivial minimum at large $b$ becomes the 
global one, meaning that chiral symmetry is broken again (dotted line). 
The corresponding behavior of the gap parameter which
minimizes the thermodynamic potential is shown in the middle
of Fig.~\ref{fntmju}.
This obviously unphysical result is a consequence of two facts: 
First, for $\mu>m_1^R(b_c)$
there are two CCMPs, i.e., twice as many ``degrees of freedom'' which 
contribute to the pressure in the non-trivial minimum. 
Second, according to Eq.~(\ref{ferm}), the Fermi momentum is lowered by the 
real part but enhanced by the imaginary part of $m_k$, and eventually becomes 
even larger than $\mu$.
These effects lead to a further enhancement of the density and thus the 
thermodynamic potential decreases faster in the non-trivial minimum than
in the trivial one.
This is also underlined by the right plot, where the pressure is displayed
as a function of the chemical potential:  
In the second chirally broken phase, the immense rise in the 
density causes the EoS even to overshoot 
the pressure of a free quark gas.

The second breaking of chiral symmetry at high chemical potential is
reminiscent to what we have found at high temperature, when we restricted
ourselves to a finite number of poles, cf.~Figs.~\ref{btmp} and
\ref{bphitmp}.
In fact, at $T=0$, the chemical potential imposes a sharp 
cutoff on the thermal part of the thermodynamic potential, and in this
sense, there is always only a finite number of poles which contribute.
For instance, the highest value of $\mu$ considered considered in the 
left plot of Fig.~\ref{fntmju} is still smaller than $m_2^R$ for
the shown range of $b$, and therefore the thermodynamics is determined
completely by the first quartet.

\section{Summary and conclusions}

A realistic continuum effective theory of strongly interacting quarks and 
gluons should incorporate the two most striking features of low-energy QCD: 
chiral symmetry breaking and confinement.
The first one represents a physically familiar concept, and it is rather 
easily modeled. 
The latter is still lacking a proper explanation, but can nevertheless be
realized by various confining criteria like, e.g., positivity violation.

The simplest possible model of chiral quark dynamics is the 
Nambu--Jona-Lasinio model, exhibiting real mass poles,
a feature shared with its nonlocal, but instantaneous generalizations.
Although very successful, these models do not incorporate confinement.
Models in which the quark propagator is an entire function represent the
opposite situation that quasiparticle poles for quarks (and gluons) are 
absent, but therefore do not allow for deconfinement.
In between, there is a large class of models that have poles in the
complex plane, which, due to the their manifest covariance,
come in complex conjugate pairs.

In this paper we have reported the observation that the thermodynamical state
in such a class of models is unstable, due to the possibility
of oscillating, or even negative, pressure. 
Further consequences can also be entropy decrease with the temperature
or
negative heat capacity, violating the standard stability criteria
for thermodynamical equilibrium.
This underlines that quark confinement is a multifaceted phenomenon that 
cannot exclusively be modeled as strong momentum dependence of dynamical mass 
and wave function renormalization. 
In this context we have investigated to what extent a suppression of 
these unphysical instabilities can be achieved by coupling the system to the 
Polyakov loop.

A simple quark model which reproduces the CCMP form of the quark 
propagator 
dynamically is a DSE model with a covariant separable interaction.
As an example, we solved such a model for a Gaussian formfactor ansatz 
at finite temperature and chemical potential
in mean-field approximation
in order to demonstrate the possible effects of CCMPs on the thermodynamics. 
The results show that CCMPs are indeed the mechanism responsible for the 
instability.
When coupling the quark propagator to the PL,
we find that the pressure instabilities are strongly reduced.
Unfortunately, more or less in the same region of the phase diagram, 
the PL itself becomes negative,
which is in sharp contrast to its standard interpretation
as exponential of the free energy of a static color source.
Thus, although there is no exact one-to-one correspondence, it seems that
one unphysical effect could only be suppressed to the expense of
a new one.

At zero temperature and finite chemical potential the PL is irrelevant.
For very high chemical potentials ($\sim 500$ MeV in our case)
CCMPs produce yet another unexpected and probably unphysical behavior: 
the chiral symmetry gets broken again.\footnote{At very high chemical 
potential chiral symmetry is expected to be broken again in a color-flavor
locked phase~\cite{Alford:1998mk}, but this is a completely different 
mechanism.}
We expect, that the same behavior is persistent in a more realistic 
setup, i.e., when one includes the effects of wave function renormalization 
as, e.g.,
in \cite{Contrera:2007wu,Hell:2008cc,Horvatic:2010md}.

From a wider perspective, vacuum instabilities in a mean-field (or 
``classical'') description of QCD are not uncommon; a prominent example being 
the tachyonic mode observed in the Savvidy vacuum 
\cite{Savvidy:1977as,Nielsen:1978rm}. 
It is interesting to note that also in this case the PL acts as a 
``stabilizer'', i.e., by suppressing the original tachyonic modes 
\cite{Meisinger:1997jt,Meisinger:2002ji}.  
Unfortunately, this program is also not entirely successful as new 
unstable modes arise \cite{Meisinger:1997jt,Meisinger:2002ji}.

Apparently, the lesson to be learnt is the following: in order
to cure the thermodynamic instability problems in the 
low-temperature and low-density domain of effective ``confining'' models 
of quark (and /or gluon) matter properly one has to go beyond the 
mean-field level of description, however clever it may be designed.
Only the explicit inclusion of the physical degrees of freedom in that domain,
the hadrons as color neutral bound states of quarks and gluons, will provide
the non-negative and monotonously rising pressure in the confinement domain
of low-energy QCD. 
As first promising works in this direction we refer to \cite{Blaschke:2007np,
Hell:2008cc,Hell:2009by,Radzhabov:2010dd}
and suggest to develop the CCMP propagator class models beyond mean field.   

\section*{Acknowledgments}
S.B. acknowledges discussions with D. Horvati\' c and G. Contrera as well as 
the hospitality of the University of Wroclaw where this work was started and 
completed.
This work was supported by the Polish Ministry for Science and Higher Education
and by CompStar, a Research Networking Programme of the European Science 
foundation.
S.B. is also supported through the project No. 119-0982930-1016
of the Ministry of Science, Education and Sports of Croatia. 
D.B. acknowledges the kind hospitality during visits at the Institut f\"ur 
Kernphysik of Technische Universit\"at Darmstadt as well as support by the 
Polish National Science Center (NCN) under grant no. NN 202 231837 and by the 
Russian Fund for Basic Research under grant no. 11-02-01538-a.
M.B. thanks D.B. for his kind hospitality at the University of Rostock 
and at ECT$^*$ in Trento
where initial work for this paper was done more than a decade ago.

\begin{appendix}
\section{Gap equation in medium}
This appendix presents the quark gap equation
in the mean-field approximation, with the Matsubara
summation being analytically performed.

Minimizing the mean-field thermodynamic potential (\ref{tpotfull}),
the quark gap at $T,\mu>0$ is obtained to be
\be
b=\frac{16 D_0}{9} T \sum_{n=-\infty}^{+\infty}
\int\frac{d^3 p}{(2\pi)^3}\mathrm{tr}_c\left[
\frac{B(\tilde{p}_n^2)f_0(\tilde{p}_n^2)}
{\tilde{p}_n^2+B^2(\tilde{p}_n^2)}\right]~,
\label{qgap}
\ee
where $\tilde{p}_n$ is to be understood as a diagonal color matrix,
see (\ref{pcol}). 
The sum over Matsubara frequencies is evaluated using the standard technique 
of rewriting it as the sum over residues of a contour integral in the complex
energy plane over the analytically continued integrand function folded with the
function (\ref{qnum}) having simple poles at the PL shifted 
Matsubara frequencies.  
The Matsubara summation is thus converted into three closed contour integrals, 
similar to what was performed in Sec.~\ref{sec:EoS}. 
These are calculated by the residue theorem, giving the result
\be
b=\frac{16 D_0}{9} \left[N_c\int\frac{d^4 p}{(2\pi)^4}
\frac{B(p^2)f_0(p^2)}{p^2+B^2(p^2)}+
2
\sum_{k=1}^\infty\int\frac{d^3 p}{(2\pi)^3}
\mathrm{Re}\left\{\mathrm{Res}(\mathcal{E}_k)
\mathrm{tr}_c\left[n_{+}(\mathcal{E}_k) + 
n_{-}(\mathcal{E}_k)\right]
\right\}\right]~.
\ee
The first term is recognized as the vacuum gap equation. Residues
can easily be deduced to be
\be
\mathrm{Res}(\mathcal{E}_k)=
\frac{B(\mathbf{p}^2,-\mathcal{E}_k^2)f_0(\mathbf{p}^2,-\mathcal{E}_k^2)}
{\mathcal{D}'(\mathcal{E}_k)}~,
\label{resgap}
\ee
with $\mathcal{D}$ given by (\ref{den}) when $A=C=1$.
An expression for a color trace of the occupation numbers
\be
\mathrm{tr}_c\left[n_{\pm}(\mathcal{E}_k)\right]=
\frac{3\Phi e^{-\beta (\mathcal{E}_k \mp\mu)}
+6\bar{\Phi} e^{-2\beta (\mathcal{E}_k\mp\mu)}+
3 e^{-3\beta (\mathcal{E}_k\mp\mu)}}
{1+3\Phi e^{-\beta (\mathcal{E}_k \mp\mu)}
+3\bar{\Phi} e^{-2\beta (\mathcal{E}_k\mp\mu)}+
e^{-3\beta (\mathcal{E}_k\mp\mu)}}~,
\ee
completes the calculation.

At zero temperature the Matsubara sum is converted to an integral
which is performed in a similar fashion as the integral for the
quark number density, see Eq.~(\ref{cont}). The result reads
\be
b=\frac{16 D_0}{9} N_c \left[\int\frac{d^4 p}{(2\pi)^4}
\frac{B(p^2)f_0(p^2)}{p^2+B^2(p^2)}+2
\sum_{k=1}^\infty\int\frac{d^3 p}{(2\pi)^3}
\mathrm{Re}\left\{\mathrm{Res}(\mathcal{E}_k)\right\}
\theta(\mu-\epsilon_k)\right]~,
\ee
with $\mathrm{Res}(\mathcal{E}_k)$ given by Eq. 
(\ref{resgap}). 
\end{appendix}

\end{document}